\documentclass[page-classic]{epl2} 

\title{Thermodynamic signature of a phase transition in the pseudogap phase of  $YBa_2Cu_3O_{x}$ high-$T_C$ superconductor}
\shorttitle{Thermodynamic signature of a phase transition in the pseudogap phase of  $YBa_2Cu_3O_{x}$} 

\author{B. Leridon\inst{1} \and P. Monod\inst{1} \and D. Colson\inst{2}}

\institute{                    
  \inst{1} Laboratoire Photons Et Mati\`ere,  UPR5/CNRS,  ESPCI \\ 10 rue Vauquelin, 75231 Paris cedex 05, France\\
  \inst{2} Service de Physique de l'Etat Condens\'e, DSM/DRECAM, CEA Saclay, 91191 Gif-Sur-Yvette Cedex, France
}

\abstract{We present here high precision magnetisation measurements in polycrystalline  $YBa_2Cu_3O_{x}$ samples, with oxygen content ranging from $x=6.19$ to $x=7.00$.  By analysing the temperature derivative of the susceptibility, we found in the underdoped superconducting samples a singular point at a temperature corresponding to $T_{mag}$, the temperature below which polarised neutrons experiments have evidenced a symmetry breaking.  We believe that this is a thermodynamic indication for the existence of a phase transition in the pseudogap state of underdoped $YBa_2Cu_3O_{x}$.}

\begin{document}

\maketitle

\section{Introduction}
Understanding the theory of high-temperature superconductors remains one of the outstanding challenges in Condensed Matter Physics. One of the fascinating questions to address is the nature of the normal state in the underdoped part of the phase diagram, and in particular of the pseudogap which develops at the Fermi energy below a temperature $T^\star$ well above the critical temperature $T_C$.   
Among the various theoretical proposals, an hypothesis is to consider the pseudogap as a precursor of the superconducting gap \cite{Emery:1995}.  In this case pairing with short-range phase coherence occurs already at $T^\star$ which is therefore more a crossover regime, while other  models predict a genuine phase transition at this temperature, associated with some order parameter as for example proposed by Varma at al.  \cite{Varma:1997,Varma:1999,Varma:2006}. The latter has recently received support from polarised neutron scattering experiments  in underdoped $YBa_2Cu_3O_{7-\delta}$ (YBCO) \cite{Fauque:2006,Mook:2008} and in $HgBa_2CuO_{4+\delta}$ \cite{Li:2008}, and possibly by the observation  of a weak magneto-optic Kerr effect rotation in zero external  field \cite{Xia:2008},  although no specific heat anomaly has been reported.  In order to search for a thermodynamic signature of this symmetry breaking, we have performed high precision measurements of the magnetic susceptibility in several underdoped and optimally doped $YBa_2Cu_3O_{x}$ (YBCOx) samples. 

\section{Measurements}
We have measured the magnetisation of sixteen YBCO polycrystalline \footnote{Typical size of the crystallites is about $30 \mu m$.} samples of various oxygen content.  The different samples with their critical temperature $T_C$ whenever superconducting (determined for simplicity by the value of T for which the magnetic susceptibility $\chi =0$ at 1Tesla) are listed in Table 1.  The samples, of about typically $10^{-3}$ mol were prepared from three different batches at SPEC-CEA (Saclay, France), then compacted into right cylinders of 6mm diameter and 6mm height, and annealed under appropriate $N_2-O_2$ mixtures to obtain different oxygen contents, using a Netzsch thermobalance.  Upon decreasing the oxygen concentration in the sample and therefore the carrier concentration, the $T_C$ of the superconducting samples was varied from 90.7 K  down to  30.7 K, and three samples were  made non superconducting. 


\begin{figure}
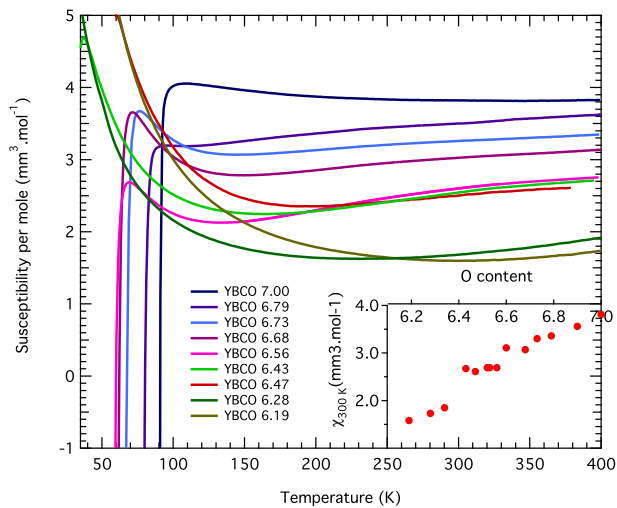

\onefigure[width=8.5cm]{Figure1.EPSF}
\caption{Susceptibility measured under 1 T as a function of temperature for a set of samples. Inset: Susceptibility measured at 300K under 1 T as a function of oxygen content x. The samples are listed in the legend from top to bottom by decreasing value of the room temperature susceptibility.}
\label{suscep1T}
\end{figure}

The samples were held in a long polyethylene straw of slightly smaller diameter and  placed into a  SQUID magnetometer, so that no sample holder correction was necessary.  Their dc magnetisation was then measured using a 6 cm-scan technique from below  $T_C$ to 400 K, under a magnetic field of 1T.
 In order to achieve the desired accuracy (besides choosing the maximum volume for the samples), the average of  nine measurements was taken at each temperature with either 1K or 2K interval steps \footnote{A special check was performed on a $x=7.00$ disc in order to measure the thermal diffusivity within the sample using the photothermal "mirage effect". This yielded a time constant of less than 10 s for our samples, compatible with the average rate of warming of $6. 10^{-3} K.s^{-1}$.  
}.  In general, the reproducibility over 5 to 10 runs of the magnetisation measurements over warming or cooling could reach $10^{-7} $emu or equivalently $10^{-8} \mu_B$ per copper atom, however some constant shift of the magnetisation (of about $10^{-6} $emu) of unknown origin occurred at times, without affecting the \textit{temperature derivative} of the magnetisation, on which we focus in this work.

\section{Experimental results}
 The susceptibilities are shown on fig. \ref{suscep1T} for a selected set of samples.
For  the nearly optimally doped samples (x=7.00), the susceptibility above $T_C$ is almost constant in temperature, evocative of a  Pauli  contribution.  However,  a small Curie contribution attributable to parasitic phases is present, which we will discuss later.

As shown in fig. \ref{suscep1T} the underdoped non superconducting samples exhibit a susceptibility first decreasing with temperature, corresponding to a Curie contribution and then increasing at high temperature. 

All the underdoped \textit{superconducting} samples exhibit a negative susceptibility  increasing at very low temperature,  corresponding to the diamagnetism of the superconducting state, then decreasing due to some Curie contribution and then increasing again at high temperatures.  The Curie term is found to vary slightly from batch to batch and to increase with underdoping. It is consistent with early observation from Johnston and co-workers \cite{Tranquada:1988,Johnston:1988}, where it is ascribed to the presence of free Cu ions.  The (uncorrected) susceptibility at 300K decreases with decreasing oxygen content, due to the loss of carriers (See the inset of Fig. \ref{suscep1T}) \cite{Tranquada:1988,Johnston:1988}.

\begin{figure}
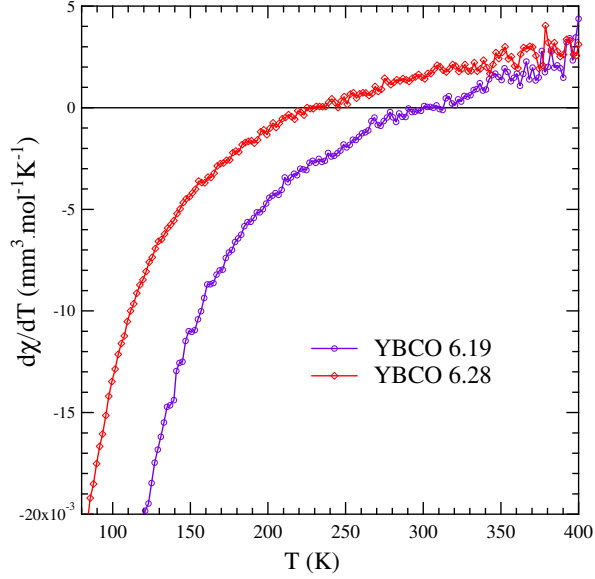

\onefigure[width=8.5cm]{Figure2.EPSF}
\caption{Derivative of the susceptibility of two different non superconducting samples with respective oxygen contents $x=6.19$ (blue circles) and $x=6.28$ (red diamonds).}
\label{dŽrivŽesAF}
\end{figure}

\begin{figure}
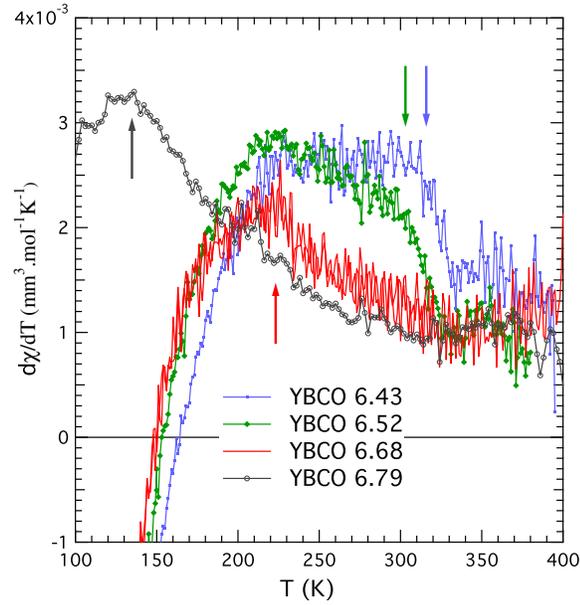

\onefigure[width=8.5cm]{Figure3.EPSF}
\caption{Derivative of the susceptibility of four different underdoped superconductive samples with oxygen contents $x=6.43$ (blue squares), $x=6.52$ (green diamonds), $x=6.68$(red line,two runs) and $x=6.79$ (grey circles). The singular points at $T_1$ are indicated by the arrows.}
\label{dŽrivŽesUD}
\end{figure}

\begin{figure}
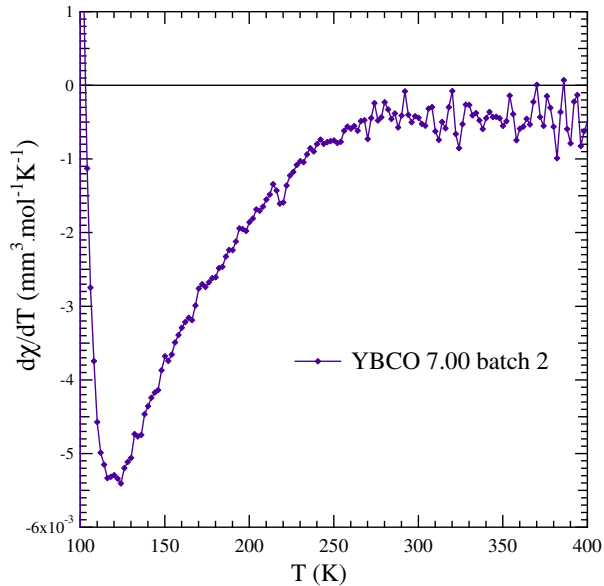

\onefigure[width=8.5cm]{Figure4.EPSF}
\caption{Derivative of the susceptibility of an optimally doped sample YBCO 7.00 (batch2).}
\label{dŽrivŽesOD}
\end{figure}

The level of accuracy of the measurements ($10^{-7} emu$ for a signal of $10^{-3} emu$ at 1 T) allows to perform a more detailed analysis.
 In order to search for any accident on these apparently smoothly varying curves, we have taken the derivative of these curves with respect to temperature and the result is shown for a set of samples in  figs. \ref{dŽrivŽesAF}, \ref{dŽrivŽesUD} and \ref{dŽrivŽesOD}.  The derivative is taken as the $\frac{\Delta M}{\Delta T}$ variation over an interval of either 1, 2 or 3K, and is centred at the midpoint.

For the non-superconducting samples (YBCO6.19, YBCO6.28 and YBCO6.34), the derivative of the susceptibility continuously increases with no singular point in our range of temperature. (See for example samples YBCO6.19 and YBCO 6.28 in fig.\ref{dŽrivŽesAF}).
However, we discovered a singular point  in the derivative of the susceptibility in every \textit{underdoped superconducting} sample (for $6.43 < x <6.79$ ) at a given temperature $T_1(x)$, x being the oxygen content.  This is illustrated in fig. \ref{dŽrivŽesUD} for sample YBCO6.43, YBCO6.52, YBCO6.68 and  YBCO6.79.  The arrows indicate the position of $T_1$. Notice that all the derivatives merge above $T_1$.

\begin{figure}
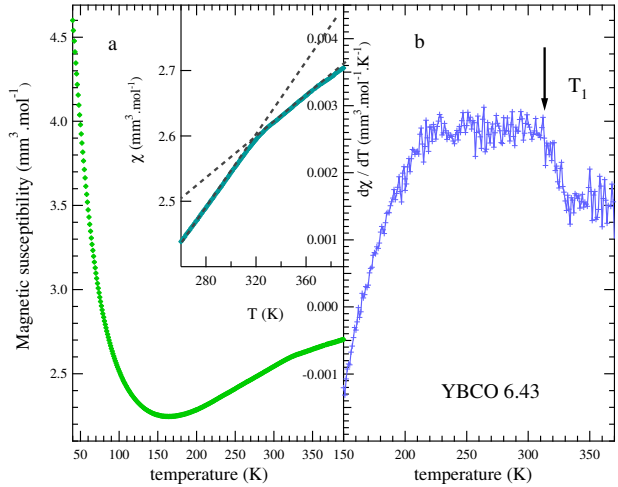

\onefigure[width=8.5cm]{Figure5.EPSF}
\caption{ a) Susceptibility of sample YBCO6.43 as a function of  temperature. Inset: enlarged view of the experimental data between 260K and 390 K. The grey dashed lines are a guide for the eye. b) Derivative of the susceptibility. The singular point in the derivative is for $T_1= 312 K \pm 2K$, and is at slightly lower temperature than the intersection of the grey  lines in 4a.}
\label{643}
\end{figure}

This singular point consists in a change in the slope of the susceptibility  at $T_1$ which is well visible on the susceptibility  of the more underdoped samples as for example YBCO6.43  at about $312 K  \pm 2$K (see fig. \ref{643}a and inset).  This translates into a downward step in the derivative of the susceptibility, whose height is found to be  of the order of $10^{-3}$ mm$^3$.mol$^{-1}$.K$^{-1}$ (see fig.\ref{643}b).  For less underdoped samples, the feature is somewhat less sharp but the anomaly is still well visible in the derivative. We choose, as a criterion for $T_1$,  the onset temperature at which the \textit{derivative} of the susceptibility starts decreasing yielding a cusp-like feature. $T_1$ is indicated by the arrows on fig. \ref{dŽrivŽesUD} and \ref{643}.  The combination of the upward trend of the Curie term and the downward trend of the normal state susceptibility produces an inflexion point in the magnetisation, yielding a wide maximum in the derivative $\frac{dM}{dT}$. $T_1$ does not  correspond necessarily to the position of this maximum which is strongly sensitive to the amplitude of the Curie term. (See for example sample 6.52 on fig. \ref{dŽrivŽesUD}.)  It may happen that the cusp and the maximum coincide as in sample YBCO 6.68. (See fig. \ref{dŽrivŽesUD}). At temperatures larger than $T_1$, for all the underdoped samples, the susceptibilities vary slowly with a similar slope,  up to our maximum temperature of 400 K.  
The susceptibilities of the four nearly optimally doped samples YBCO7b1 , YBCO7b2, YBCO7b3 ($x=7.00$) and  YBCO6.90 do not exhibit such a singular point at least within our level of accuracy ($\pm10^{-7}$emu.K$^{-1}$). (See sample YBCO7b2 in fig. \ref{dŽrivŽesOD}.)  

 \begin{table}
 \caption{Summary of the results for different samples. 
}
\begin{tabular}[width=8.5cm]{c | c | c | c | c | c | c }

 \\
Sample  & Oxygen & $T_{C} $(K) & $T_{1} (K)$ & $\chi_{300K}$ & $\frac{T}{\chi} \frac{d\chi}{dT}$ & batch  \\ 
name & content & & & (uncorr.) &  at $T_1$ & \# \\
 \hline
& $ \pm 0.01$& $\pm$ 1 K && $mm^3.$ & & \\
&&&&$mol^{-1}$&&\\
YBCO6.19 & 6.19 & - & - & 1.58 & -& 3\\
YBCO6.28 & 6.28 & - & - &  1.73 & -&3 \\
YBCO6.34 & 6.34 & -  & - &  1.85 & -&3 \\
YBCO6.43 & 6.43 & 30.7 & 312 $\pm 2K$&  2.67  & 0.31 & 3  \\
 YBCO6.47 & 6.47 & 45 &  317 $\pm 4K$ & 2.61 & 0.22 &1 \\  
 YBCO6.52 & 6.52 &  55.1 &  302 $\pm 2K $&  2.69    & 0.27 & 3 \\
  YBCO6.53 & 6.53 & 57.5 &  315 $\pm 5K$ & 2.69 & 0.28 & 2\\  
 YBCO6.56 & 6.56 & 59.8 &  312  $\pm 2K$ & 2.69 &  0.26 & 2\\    
 YBCO6.60  & 6.60& 60.8 &  238 $\pm6K$ & 3.11 & 0.19 & 1\\  
YBCO6.68 & 6.68 & 62.2 & 226 $\pm2K$ & 3.07 & 0.18 &  1 \\   
 YBCO6.73 & 6.73 & 67.5 &  212 $\pm 6K$ & 3.30 & 0.13 &  1\\   
  YBCO6.79 & 6.79 & 80.0 &  133  $\pm 6K$& 3.36 &  0.15 &  3\\   
  YBCO6.90 & 6.90 & 90.9 &- & 3.56 & -& 3 \\ 
YBCO7b1& 7.00 & 90.7 & - & 3.95 & -&  1\\  
YBCO7b2& 7.00 & 90.7 &  - & 3.94 & -&  2\\  
YBCO7b3& 7.00 & 90.6 &  - & 3.81  & -&  3\\  
 
 \end{tabular}

 \end{table}

\section{Analysis and discussion}
In view of the small size of this effect, it is essential to check that this singular point is not related to the presence of parasitic phases at small concentration that are always present.
The $Y_2BaCuO_5$ "green phase" was therefore checked quantitatively with EPR identification.  This yielded a level of about only $0.1\%$ mole, which is not consistent with the amplitude of the Curie term found in the optimally doped samples. This Curie term is indeed consistent with the presence of  a few part per thousands of $BaCuO_2$(namely $ Ba_{48}Cu_{48}0_{96}$) for which the apparent ferromagnetic  Weiss  temperature is  of $+40 K$ in the range 150-300 K, as observed here and whose susceptibility is known to be featureless in this range \cite{Wang:1995}.
The  presence  of $CuO$ was  checked  by  XRD to  be less  than  $1\%$, thus making impossible for us the detection of the slope discontinuity of the susceptibility at $T_N^{CuO}=230 K$ \cite{Laurie:1998}.  A special run carried out on a pure CuO sample that had received the same oxygen treatment as the one necessary to make an underdoped YBCO sample with $x=6.43$  did not show any measurable change of  $T_N^{CuO}$. Therefore none of these well characterised phases can account for the anomalies in the derivative of the susceptibility at $T_1$ \footnote{ It was also observed, for all the samples issued from batch 3, that at low magnetic field a well defined, irreversible in temperature, magnetisation step-like feature of variable amplitude (typically $10^{-6}$ emu ) occurred at 338K. This very small contribution is attributed to the presence of an hitherto unspecified magnetic phase extrinsic to YBCO. Since only batch 3 is concerned, this phase can not be responsible for the anomalies at $T_1$.}.

Finally, systematic measurements of the magnetisation as a function of field  in the full range of field and temperature available yielded  an overwhelming linear field dependance superimposed over a small S-shaped hysteresis cycle saturating above 2 kgauss. This cycle, observed in all batches has a width of 300 gauss at 150 K and its saturation magnetisation decreases by 10\% between 300K and 400 K. The amplitude of the magnetic moment associated to this cycle, for which we have no definite interpretation, is about $10^{-2} $emu.mol$^{-1}$. Such cycles were previously reported \cite{Panagopoulos:2006} but the variation in the magnetic moment amplitude observed from batch to batch rules out any intrinsic origin. More recently, Xia et al. \cite{Xia:2008} have observed through polar Kerr effect measurements a rotation of the polarisation of light attributed to a magnetic moment below a temperature $T_S$. We have performed the same field training procedure as the authors of ref \cite{Xia:2008}.  We observe on sample YBCO6.60 a difference of magnetisation measured under 100 gauss between the $+3 $T and the $-3$T preparation of about $10^{-7} \mu_B/$Cu. This difference of magnetisation is observed to be temperature independent between 100K and 300K  at better than $10^{-8} \mu_B/$Cu. In similar samples, Xia et al \cite{Xia:2008} find  $T_S\sim160 $K, which means that their signal corresponds to a magnetisation of less than $10^{-8} \mu_B$ per YBCO copper atom, in order to be consistent with our data. 

\begin{figure}
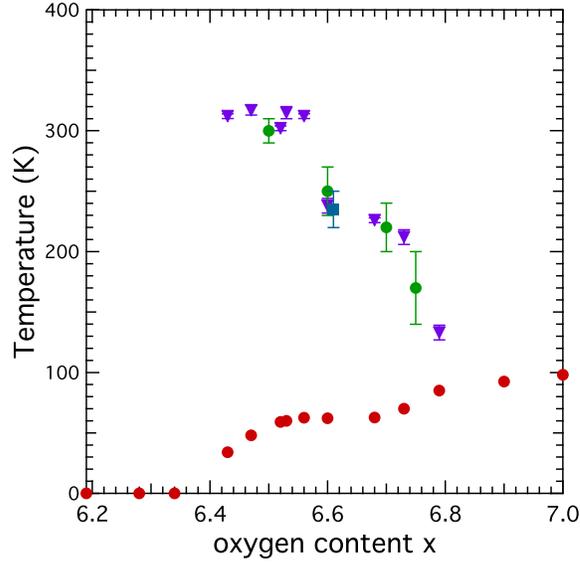

\begin{center}
\onefigure[width=8.5cm]{Figure6.EPSF}
\caption{ Phase diagram of the underdoped YBCO samples. Red dots: Critical temperature $T_C$; down blue triangles $T_1$ deduced from the position of the non-analytic point in the derivative; Green dots with error bars $T_{mag}$ after Fauqu\'e et al.\cite{Fauque:2006}; blue square with error bar, $T_{mag}$ after Mook et al \cite{Mook:2008}.}
\label{phadiag}
\end{center}
\end{figure}

The temperature  $T_1$ is plotted on fig.  \ref{phadiag} (blue down triangles) as well as $T_C$ (red dots) and the temperatures $T_{mag}$ obtained by Fauqu\'e and co-workers \cite{Fauque:2006} (green dots) and by Mook and co-workers \cite{Mook:2008} (blue diamond).
$T_1$ is found to increase with underdoping and shows a remarkable agreement with $T_{mag}$.
This seems to establish a relationship between the anomalous character of the thermodynamic quantity $\frac{d\chi}{dT}$ and the symmetry breaking evidenced by polarised neutron experiments.

In order to establish a relationship between the uniform magnetisation measured in this experiment and the staggered magnetisation evidenced by Fauqu\'e \cite{Fauque:2006}, one way is to assume that the staggered magnetisation is coupled in second order to the uniform magnetisation in a way similar to that of an antiferromagnet.
A mean field calculation  \cite{Gronsleth:2008} then gives an estimate for the quantity $\frac{T_1}{\chi}\frac{d\chi}{dT}\sim1$ just below $T_1$ in the presence of staggered magnetisation.   The corresponding value measured in the experiments is typically 0.15-0.3 in dimensionless units (see table 1), which we consider as a rather satisfactory agreement in view of the fact that the contribution of the (uncorrected) Curie like terms gives a negative value for this quantity.

No evidence for a discontinuity in the specific heat in the underdoped regime of YBCO  has  been reported to date in the literature, however the observed magnetic moments \cite{Fauque:2006,Mook:2008} are of the order $0.1 \mu_B/$Cu, which should lead to a sizeable change of entropy.   In this respect, one should note that in the case of the antiferromagnetic transition of $La_2CuO_4$ at 240 K, Sun et al. \cite{Sun:1990} have found no evidence for a  specific anomaly at the transition in spite of a spectacular increase of the susceptibility, due to the very weak entropy variation involved in the 3D spin ordering. A more fundamental explanation might be that the staggered magnetic moments evidenced by Fauqu\'e  and co-workers \cite{Fauque:2006} are predicted to be described by an Ashkin-Teller model \cite{Baxter:1982,Aji:2007,Borkje:2008,Gronsleth:2008}.  For a range of parameters in such a model, the specific heat is not expected to have any singular behaviour, although the susceptibility to an external magnetic field is singular \cite{Baxter:1982}. 
Since our present work  is based on  ceramic samples, a search for the corresponding (and possibly anisotropic) susceptibility anomalies in single crystals appears as a further test of these results \footnote{For the crystals used in Ref. \cite{Xia:2008}, the signal would be about two orders of magnitude smaller than in our samples, therefore below the noise level of our system. On the other hand, large single crystals usually contain a substantial amount of green phase producing a strong Curie term and thus making it difficult the observation of such small effects.}.

\section{Conclusion}
The present evidence of an apparently non-analytic variation of the temperature derivative of the susceptibility of YBCO within the pseudogap temperature range appears in direct contradiction with an essentially "featureless" crossover regime predicted by several models \cite{Emery:1995}.  The overall agreement for the temperature of this susceptibility anomaly with that of the onset of staggered magnetisation points out toward a phase transition whose order parameter is weakly coupled to the external magnetic field.

\acknowledgments
We gratefully acknowledge the following persons for stimulating discussions or experimental collaboration:  V. Aji, Ph. Bourges, B. Fauqu\'e, D. Fournier, A. Kapitulnik, W. Rischau, A. Shekhter, Ch. Simon and C.M. Varma.


\begin{thebibliography}{18}
\expandafter\ifx\csname natexlab\endcsname\relax\def\natexlab#1{#1}\fi
\expandafter\ifx\csname bibnamefont\endcsname\relax
  \def\bibnamefont#1{#1}\fi
\expandafter\ifx\csname bibfnamefont\endcsname\relax
  \def\bibfnamefont#1{#1}\fi
\expandafter\ifx\csname citenamefont\endcsname\relax
  \def\citenamefont#1{#1}\fi
\expandafter\ifx\csname url\endcsname\relax
  \def\url#1{\texttt{#1}}\fi
\expandafter\ifx\csname urlprefix\endcsname\relax\def\urlprefix{URL }\fi
\providecommand{\bibinfo}[2]{#2}
\providecommand{\eprint}[2][]{\url{#2}}

\bibitem[{\citenamefont{Emery and Kivelson}(1995)}]{Emery:1995}
\bibinfo{author}{\bibfnamefont{V.}~\bibnamefont{Emery}} \bibnamefont{and}
  \bibinfo{author}{\bibfnamefont{S.}~\bibnamefont{Kivelson}},
  \bibinfo{journal}{Nature} \textbf{\bibinfo{volume}{374}},
  \bibinfo{pages}{434} (\bibinfo{year}{1995}).

\bibitem[{\citenamefont{Varma}(1997)}]{Varma:1997}
\bibinfo{author}{\bibfnamefont{C.}~\bibnamefont{Varma}},
  \bibinfo{journal}{Phys. Rev. B} \textbf{\bibinfo{volume}{55}},
  \bibinfo{pages}{14554} (\bibinfo{year}{1997}).

\bibitem[{\citenamefont{Varma}(1999)}]{Varma:1999}
\bibinfo{author}{\bibfnamefont{C.}~\bibnamefont{Varma}},
  \bibinfo{journal}{Phys. Rev. Letters} \textbf{\bibinfo{volume}{83}},
  \bibinfo{pages}{3538} (\bibinfo{year}{1999}).

\bibitem[{\citenamefont{Varma}(2006)}]{Varma:2006}
\bibinfo{author}{\bibfnamefont{C.}~\bibnamefont{Varma}},
  \bibinfo{journal}{Phys. Rev. B} \textbf{\bibinfo{volume}{73}},
  \bibinfo{pages}{155113} (\bibinfo{year}{2006}).

\bibitem[{\citenamefont{Fauque et~al.}(2006)\citenamefont{Fauque, Sidis,
  Hinkov, Pailhes, Lin, Chaud, and Bourges}}]{Fauque:2006}
\bibinfo{author}{\bibfnamefont{B.}~\bibnamefont{Fauque}},
  \bibinfo{author}{\bibfnamefont{Y.}~\bibnamefont{Sidis}},
  \bibinfo{author}{\bibfnamefont{V.}~\bibnamefont{Hinkov}},
  \bibinfo{author}{\bibfnamefont{S.}~\bibnamefont{Pailhes}},
  \bibinfo{author}{\bibfnamefont{C.~T.} \bibnamefont{Lin}},
  \bibinfo{author}{\bibfnamefont{X.}~\bibnamefont{Chaud}}, \bibnamefont{and}
  \bibinfo{author}{\bibfnamefont{P.}~\bibnamefont{Bourges}},
  \bibinfo{journal}{Phys. Rev.Letters} \textbf{\bibinfo{volume}{96}},
  \bibinfo{eid}{197001} (pages~\bibinfo{numpages}{4}) (\bibinfo{year}{2006}),
  \urlprefix\url{http://link.aps.org/abstract/PRL/v96/e197001}.

\bibitem[{\citenamefont{Mook et~al.}(2008)\citenamefont{Mook, Sidis, Fauqu\'e,
  Bal\'edent, and Bourges}}]{Mook:2008}
\bibinfo{author}{\bibfnamefont{H.~A.} \bibnamefont{Mook}},
  \bibinfo{author}{\bibfnamefont{Y.}~\bibnamefont{Sidis}},
  \bibinfo{author}{\bibfnamefont{B.}~\bibnamefont{Fauqu\'e}},
  \bibinfo{author}{\bibfnamefont{V.}~\bibnamefont{Bal\'edent}},
  \bibnamefont{and} \bibinfo{author}{\bibfnamefont{P.}~\bibnamefont{Bourges}},
  \bibinfo{journal}{Phys. Rev. B} \textbf{\bibinfo{volume}{78}},
  \bibinfo{pages}{020506} (\bibinfo{year}{2008}).

\bibitem[{\citenamefont{Li et~al.}(2008)\citenamefont{Li, Baledent, braisic,
  Cho, Fauque, Sidis, Yu, Zhao, Bourges, and Greven}}]{Li:2008}
\bibinfo{author}{\bibfnamefont{Y.}~\bibnamefont{Li}},
  \bibinfo{author}{\bibfnamefont{V.}~\bibnamefont{Baledent}},
  \bibinfo{author}{\bibfnamefont{N.}~\bibnamefont{braisic}},
  \bibinfo{author}{\bibfnamefont{Y.}~\bibnamefont{Cho}},
  \bibinfo{author}{\bibfnamefont{B.}~\bibnamefont{Fauque}},
  \bibinfo{author}{\bibfnamefont{Y.}~\bibnamefont{Sidis}},
  \bibinfo{author}{\bibfnamefont{G.}~\bibnamefont{Yu}},
  \bibinfo{author}{\bibfnamefont{X.}~\bibnamefont{Zhao}},
  \bibinfo{author}{\bibfnamefont{P.}~\bibnamefont{Bourges}}, \bibnamefont{and}
  \bibinfo{author}{\bibfnamefont{M.}~\bibnamefont{Greven}},
  \bibinfo{journal}{Nature} \textbf{\bibinfo{volume}{455}},
  \bibinfo{pages}{372} (\bibinfo{year}{2008}).

\bibitem[{\citenamefont{Xia et~al.}(2008)\citenamefont{Xia, Schemm, Deutscher,
  Kivelson, Bonn, Hardy, Liang, Siemons, Koster, Fejer et~al.}}]{Xia:2008}
\bibinfo{author}{\bibfnamefont{J.}~\bibnamefont{Xia}},
  \bibinfo{author}{\bibfnamefont{E.}~\bibnamefont{Schemm}},
  \bibinfo{author}{\bibfnamefont{G.}~\bibnamefont{Deutscher}},
  \bibinfo{author}{\bibfnamefont{S.~A.} \bibnamefont{Kivelson}},
  \bibinfo{author}{\bibfnamefont{D.~A.} \bibnamefont{Bonn}},
  \bibinfo{author}{\bibfnamefont{W.~N.} \bibnamefont{Hardy}},
  \bibinfo{author}{\bibfnamefont{R.}~\bibnamefont{Liang}},
  \bibinfo{author}{\bibfnamefont{W.}~\bibnamefont{Siemons}},
  \bibinfo{author}{\bibfnamefont{G.}~\bibnamefont{Koster}},
  \bibinfo{author}{\bibfnamefont{M.~M.} \bibnamefont{Fejer}},
  \bibnamefont{et~al.}, \bibinfo{journal}{Phys. Rev. Lett.}
  \textbf{\bibinfo{volume}{100}}, \bibinfo{pages}{127002}
  (\bibinfo{year}{2008}).

\bibitem[{\citenamefont{Tranquada et~al.}(1988)\citenamefont{Tranquada,
  Moudden, Goldman, Zolliker, Cox, Shirane, Sihha, Vaknin, Johnston, Alvarez
  et~al.}}]{Tranquada:1988}
\bibinfo{author}{\bibfnamefont{J.}~\bibnamefont{Tranquada}},
  \bibinfo{author}{\bibfnamefont{A.}~\bibnamefont{Moudden}},
  \bibinfo{author}{\bibfnamefont{A.}~\bibnamefont{Goldman}},
  \bibinfo{author}{\bibfnamefont{P.}~\bibnamefont{Zolliker}},
  \bibinfo{author}{\bibfnamefont{D.}~\bibnamefont{Cox}},
  \bibinfo{author}{\bibfnamefont{G.}~\bibnamefont{Shirane}},
  \bibinfo{author}{\bibfnamefont{S.}~\bibnamefont{Sihha}},
  \bibinfo{author}{\bibfnamefont{D.}~\bibnamefont{Vaknin}},
  \bibinfo{author}{\bibfnamefont{D.}~\bibnamefont{Johnston}},
  \bibinfo{author}{\bibfnamefont{M.~S.} \bibnamefont{Alvarez}},
  \bibnamefont{et~al.}, \bibinfo{journal}{Phys. Rev. B}
  \textbf{\bibinfo{volume}{38}}, \bibinfo{pages}{2477} (\bibinfo{year}{1988}).

\bibitem[{\citenamefont{Johnston et~al.}(1988)\citenamefont{Johnston, Sinha,
  Jacobson, and Newsam}}]{Johnston:1988}
\bibinfo{author}{\bibfnamefont{D.}~\bibnamefont{Johnston}},
  \bibinfo{author}{\bibfnamefont{S.}~\bibnamefont{Sinha}},
  \bibinfo{author}{\bibfnamefont{A.}~\bibnamefont{Jacobson}}, \bibnamefont{and}
  \bibinfo{author}{\bibfnamefont{J.}~\bibnamefont{Newsam}},
  \bibinfo{journal}{Physica C} \textbf{\bibinfo{volume}{153}},
  \bibinfo{pages}{572} (\bibinfo{year}{1988}).

\bibitem[{\citenamefont{Wand et~al.}(1995)\citenamefont{Wand, Jonhston, Miller,
  and Vaknin}}]{Wang:1995}
\bibinfo{author}{\bibfnamefont{Z.}~\bibnamefont{Wand}},
  \bibinfo{author}{\bibfnamefont{D.}~\bibnamefont{Jonhston}},
  \bibinfo{author}{\bibfnamefont{L.}~\bibnamefont{Miller}}, \bibnamefont{and}
  \bibinfo{author}{\bibfnamefont{D.}~\bibnamefont{Vaknin}},
  \bibinfo{journal}{Phys. Rev. B} \textbf{\bibinfo{volume}{52}},
  \bibinfo{pages}{10} (\bibinfo{year}{1995}).

\bibitem[{\citenamefont{Laurie et~al.}(1998)\citenamefont{Laurie, Franck, and
  Lin}}]{Laurie:1998}
\bibinfo{author}{\bibfnamefont{D.}~\bibnamefont{Laurie}},
  \bibinfo{author}{\bibfnamefont{J.}~\bibnamefont{Franck}}, \bibnamefont{and}
  \bibinfo{author}{\bibfnamefont{C.-T.} \bibnamefont{Lin}},
  \bibinfo{journal}{Physica C} \textbf{\bibinfo{volume}{297}},
  \bibinfo{pages}{59} (\bibinfo{year}{1998}).

\bibitem[{\citenamefont{Panagopoulos et~al.}(2006)\citenamefont{Panagopoulos,
  Majoros, Nishizaki, and Iwasaki}}]{Panagopoulos:2006}
\bibinfo{author}{\bibfnamefont{C.}~\bibnamefont{Panagopoulos}},
  \bibinfo{author}{\bibfnamefont{M.}~\bibnamefont{Majoros}},
  \bibinfo{author}{\bibfnamefont{T.}~\bibnamefont{Nishizaki}},
  \bibnamefont{and} \bibinfo{author}{\bibfnamefont{H.}~\bibnamefont{Iwasaki}},
  \bibinfo{journal}{Phys. Rev. Lett.} \textbf{\bibinfo{volume}{96}},
  \bibinfo{pages}{047002} (\bibinfo{year}{2006}).

\bibitem[{\citenamefont{Gronsleth et~al.}(2008)\citenamefont{Gronsleth,
  Nilssen, Dahl, Varma, and Sudb\o}}]{Gronsleth:2008}
\bibinfo{author}{\bibfnamefont{M.~S.} \bibnamefont{Gronsleth}},
  \bibinfo{author}{\bibfnamefont{T.~B.} \bibnamefont{Nilssen}},
  \bibinfo{author}{\bibfnamefont{E.~K.} \bibnamefont{Dahl}},
  \bibinfo{author}{\bibfnamefont{C.~M.} \bibnamefont{Varma}}, \bibnamefont{and}
  \bibinfo{author}{\bibfnamefont{A.}~\bibnamefont{Sudb\o}},
  \bibinfo{journal}{http://arxiv.org/abs/0806.2665}  (\bibinfo{year}{2008}).

\bibitem[{\citenamefont{Sun et~al.}(1990)\citenamefont{Sun, Cho, Chou, Lee,
  Miller, Johnston, Hidaka, and Murakami}}]{Sun:1990}
\bibinfo{author}{\bibfnamefont{K.}~\bibnamefont{Sun}},
  \bibinfo{author}{\bibfnamefont{J.}~\bibnamefont{Cho}},
  \bibinfo{author}{\bibfnamefont{F.}~\bibnamefont{Chou}},
  \bibinfo{author}{\bibfnamefont{W.}~\bibnamefont{Lee}},
  \bibinfo{author}{\bibfnamefont{L.}~\bibnamefont{Miller}},
  \bibinfo{author}{\bibfnamefont{D.}~\bibnamefont{Johnston}},
  \bibinfo{author}{\bibfnamefont{Y.}~\bibnamefont{Hidaka}}, \bibnamefont{and}
  \bibinfo{author}{\bibfnamefont{T.}~\bibnamefont{Murakami}},
  \bibinfo{journal}{Phys. Rev. B} \textbf{\bibinfo{volume}{43}},
  \bibinfo{pages}{239} (\bibinfo{year}{1990}).

\bibitem[{\citenamefont{Baxter}(1982)}]{Baxter:1982}
\bibinfo{author}{\bibfnamefont{R.}~\bibnamefont{Baxter}},
  \emph{\bibinfo{title}{{Exactly Solved Models in Statistical Mechanics}}}
  (\bibinfo{publisher}{Academic Press}, \bibinfo{year}{1982}).

\bibitem[{\citenamefont{Aji and Varma}(2007)}]{Aji:2007}
\bibinfo{author}{\bibfnamefont{V.}~\bibnamefont{Aji}} \bibnamefont{and}
  \bibinfo{author}{\bibfnamefont{C.~M.} \bibnamefont{Varma}},
  \bibinfo{journal}{Phys. Rev. Lett.} \textbf{\bibinfo{volume}{99}},
  \bibinfo{pages}{167003} (\bibinfo{year}{2007}).

\bibitem[{\citenamefont{Borkje and Sudb\o}(2008)}]{Borkje:2008}
\bibinfo{author}{\bibfnamefont{K.}~\bibnamefont{Borkje}} \bibnamefont{and}
  \bibinfo{author}{\bibfnamefont{A.}~\bibnamefont{Sudb\o}},
  \bibinfo{journal}{Phys. Rev. B} \textbf{\bibinfo{volume}{77}},
  \bibinfo{pages}{092404} (\bibinfo{year}{2008}).

\end{thebibliography}

\end{document}